# Thermal conductivity of AlN, GaN, and Al$_x$Ga$_{1-x}$N alloys as a function of composition, temperature, crystallographic direction, and isotope disorder from first principles

Sahil Dagli, Kelsey A. Mengle, and Emmanouil Kioupakis [a]

Department of Materials Science and Engineering, University of Michigan, Ann Arbor, Michigan 48109, USA

Ultra-wide-band-gap group-III nitrides are of interest for applications in deep-ultraviolet optoelectronics and power electronics. Such devices must be able to efficiently dissipate heat generated from their operation, making the thermal conductivity of the constituent materials an important parameter for high-power applications. We have investigated the phonon-limited thermal conductivity of AlN, GaN, and Al$_x$Ga$_{1-x}$N using first-principles calculations, with a focus on the effects of compositional and isotopic disorder. Our Boltzmann-transport-equation calculations show that the maximum thermal conductivity for AlN (GaN) is 348 W m$^{-1}$ K$^{-1}$ (235 W m$^{-1}$ K$^{-1}$) for with pure $^{14}$N, and 292 W m$^{-1}$ K$^{-1}$ for GaN with pure $^{71}$Ga. Al$_x$Ga$_{1-x}$N alloys reach a minimum thermal conductivity at Al mole fractions of x = 0.60 to 0.71 over the 100-1000K temperature range. Our results provide understanding on the effects of isotope disorder on the thermal conductivity of AlN and GaN. We also present an analytical model for the evaluation of the thermal conductivity of Al$_x$Ga$_{1-x}$N alloys for arbitrary composition and temperature, which can be applied for the thermal design of AlGaN-based electronic and optoelectronic devices.

AlN, GaN, and Al$_x$Ga$_{1-x}$N find applications in high-power electronic and optoelectronic devices due to their thermal stability and ultra-wide direct band gaps.[1–3] However, self-heating during operation inhibits the performance of high-power devices.[4] Therefore, devices must be designed to effectively dissipate the generated heat, making the thermal conductivity of the constituent materials, $\kappa$, an important design parameter.

---

[a] Electronic mail: kioup@umich.edu

The thermal conductivities of wurtzite AlN, GaN, and $Al_xGa_{1-x}N$ have been investigated previously both experimentally using the $3\omega$ technique[5–7] and theoretically using the virtual crystal approximation (VCA),[8] the Callaway model,[9–11] and the relaxation time approximation (RTA) within the Boltzmann transport equation (BTE).[12,13] Reported values for the room-temperature thermal conductivity of GaN are approximately 230 W m$^{-1}$ K$^{-1}$ from experiment[5,14,15] and 239 W m$^{-1}$ K$^{-1}$ from theory,[16] while AlN values are approximately 263-285 W m$^{-1}$ K$^{-1}$ experimentally[6,17,18] and 317-319 W m$^{-1}$ K$^{-1}$ theoretically.[6,18] Theoretical studies of these two compounds typically use the Callaway model,[9,10,19] which can be a better predictor of $\kappa$ than the RTA of the BTE due to its distinguishing treatment of normal and Umklapp phonon-scattering processes;[20] however, the Callaway model cannot guarantee a systematic improvement to the RTA.[19]

In addition to the binary compounds, the thermal conductivity of $Al_xGa_{1-x}N$ alloys has also been calculated previously using a combination of the VCA and Callaway models.[8] With the advancement of computational tools, the thermal conductivity of materials can be calculated with the full BTE while investigating the various factors that affect phonon scattering. The existence of any disorder in the lattice increases phonon scattering, thus decreasing $\kappa$. This motivates the study of isotopically pure samples as a means to increase $\kappa$. Using pure Ga isotopes in GaN has been theoretically shown to increase the thermal conductivity by 65% at room temperature,[16] however, the effect of using pure N isotopes remains an open question.

In this work, we determine the phonon-limited thermal conductivity of AlN, GaN, and $Al_xGa_{1-x}N$ as a function of temperature, alloy composition, crystallographic direction, and isotope substitution, using atomistic calculations based on density functional theory (DFT) and the full BTE. We investigate the effects of different N isotope substitution in AlN and GaN, Ga isotope substitution in GaN, and alloy disorder in $Al_xGa_{1-x}N$. We fit our data for the natural isotope ratio with a consistent mathematical model that predicts the thermal conductivity of $Al_xGa_{1-x}N$ alloys at all temperatures and compositions. Our results guide experimental measurements of the thermal conductivity of $Al_xGa_{1-x}N$ alloys and help engineer the design of devices for efficient thermal management.

Our calculations were performed with the full BTE as implemented in almaBTE.[21] The input files, generated using DFT, were obtained from the online almaBTE database. The Γ-centered Brillouin-zone sampling grids were increased up to 32×32×32 to ensure convergence of the thermal conductivity to within 2%. To study the effects of isotope ratios, the ratios of $^{14}N$ to $^{15}N$ and $^{69}Ga$ to $^{71}Ga$ were manually changed in the almaBTE source code to reflect the correct average masses. Al isotope effects were not included as the standard atomic mass of Al is only determined by one isotope.[22] For calculations on $Al_xGa_{1-x}N$, the virtual crystal approximation (VCA) was used to simulate alloy behavior by calculating average atomic behavior between AlN and GaN.[21]

Figure 1(a,c) shows the thermal conductivity of AlN and GaN along the **a** direction of wurtzite with varying N isotope ratios across the temperature range from 100 to 1000 K. The thermal conductivity decreases with increasing temperature, reflecting the stronger phonon-phonon scattering at elevated temperatures. The results along the **c** direction [Fig. 1(b,d)] show similar trends. The AlN thermal conductivity at 300 K along **c** for the naturally occurring isotope ratio (99.6% $^{14}N$), $\kappa_{nat}$ (339 W m$^{-1}$ K$^{-1}$), agrees with previous theoretical results by Slack *et al.* (319 W m$^{-1}$ K$^{-1}$) to within 6%.[6] The oxygen correction included by Slack *et al.* to complement

their experimental data overestimates the thermal conductivity for temperatures below 300 K.[6] Experimental results from Rounds *et al.* agree with our calculations to within 6%.[7] Differences between our results and Xu *et al.* can be attributed to our calculations considering only defect-free materials.[18] Our GaN results for the natural isotope ratio along **c** also agree with Carrete *et al.* to within 8-10% across the temperature range 200 to 500 K.[21] Our $\kappa_{nat}$ for GaN at 300 K along **a** is also in agreement with the experimental value from Mion *et al.* of 230 W m$^{-1}$ K$^{-1}$ to within 2%.[5]

Table I lists the AlN and GaN thermal conductivities for different $^{14}$N isotope ratios at 300 K along **a** and **c**. The maximum value occurs for pure $^{14}$N and decreases as the $^{14}$N fraction decreases. AlN has a higher maximal thermal conductivity (348 W m$^{-1}$ K$^{-1}$) than GaN (235 W m$^{-1}$ K$^{-1}$). The minimum value for AlN occurs at the equimolar $^{14}$N:$^{15}$N composition, while for GaN the minimum occurs for 25% $^{14}$N. Past these compositions, the thermal conductivity increases as the $^{15}$N fraction increases. Table II lists literature values of the thermal conductivity, including experimental samples synthesized by physical vapor transport (PVT) and hydride vapor phase epitaxy (HVPE), as well as first-principles calculations and analytical corrections.

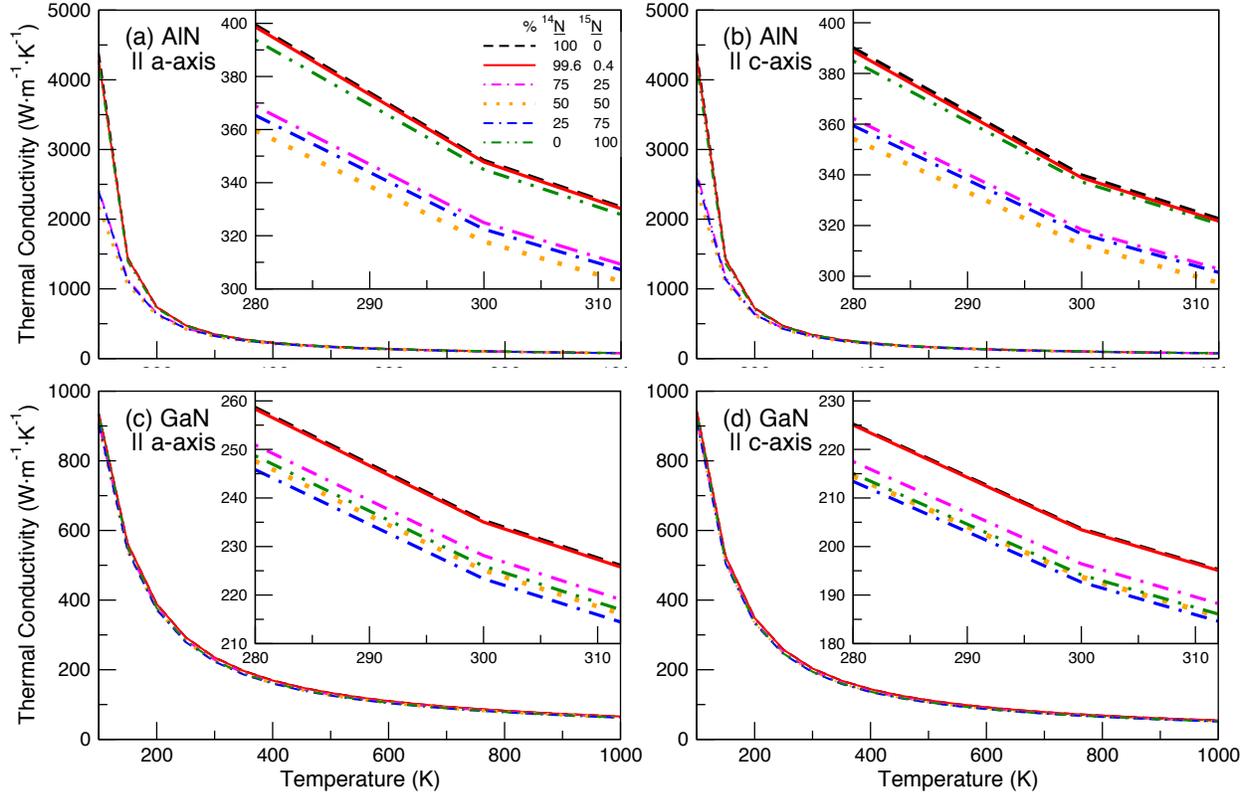

FIG 1. Thermal conductivity of AlN calculated along the (a) **a** and (b) **c**-directions and GaN calculated along the (c) **a** and (d) **c**-directions for different ratios of N isotopes over the temperature range from 100 to 1000 K. The insets highlight the region near room temperature (280 to 310 K) to illustrate the isotope effects.

TABLE I. Thermal conductivity of AlN and GaN at 300 K (in W m$^{-1}$ K$^{-1}$) along the **a** and **c** crystallographic directions of the wurtzite structure as a function of N isotope substitution.

| | AlN | |
|---|---|---|
| % $^{14}$N | **a**-direction | **c**-direction |
| 100 | 348 | 340 |
| 99.6 (natural) | 348 | 339 |
| 75 | 325 | 318 |
| 50 | 318 | 312 |
| 25 | 322 | 317 |
| 0 | 345 | 337 |

| | GaN | |
|---|---|---|
| % $^{14}$N | **a**-direction | **c**-direction |
| 100 | 235 | 204 |
| 99.6 (natural) | 235 | 203 |
| 75 | 228 | 196 |
| 50 | 225 | 194 |
| 25 | 223 | 193 |
| 0 | 226 | 194 |

TABLE II. Room temperature thermal conductivity values reported in literature for AlN and GaN (in W m$^{-1}$ K$^{-1}$).

| Methods | AlN |
|---|---|
| PVT, 3ω | 263[18] |
| Analytical model, including defects (**c**-axis) | 226[18] |
| First principles, including defects (**c**-axis) | 197[18] |
| HPVE, 3ω | 341[7] |
| PVT, 3ω | 374[7] |
| PVT, heat flow w/ analytical correction (**c**-axis) | 319[6] |
| Sputter, 3ω | 285[17] |
| This work, first principles (**a**-axis) | 348 |
| This work, first principles (**c**-axis) | 339 |
| Methods | GaN |
| HPVE, 3ω | 230[5] |
| HPVE, axial stationary heat flow (**c**-axis) | 230[14] |
| PVT, heat flow w/ analytical correction (**c**-axis) | 227[15] |
| First principles (**c**-axis) | 239[16] |
| This work, first principles (**a**-axis) | 235 |
| This work, first principles (**c**-axis) | 203 |

We also examined the effect of cation isotope substitution on the thermal conductivity. Since only trace amounts of Al isotopes other than $^{27}$Al occur, we focus on Ga isotopes. Figure 2 shows the thermal conductivity of GaN along **a** and **c** with varying Ga isotope ratios from 100 to 1000 K. Table III lists the thermal conductivity for different $^{69}$Ga isotope ratios at 300 K for GaN along both **a** and **c**. The highest thermal conductivity occurs at 0% $^{69}$Ga (100% $^{71}$Ga), with the highest value being 294 W m$^{-1}$ K$^{-1}$ along **a**. Using pure Ga isotopes increases the thermal conductivity by 25% along **a**, and 15% along **c**.

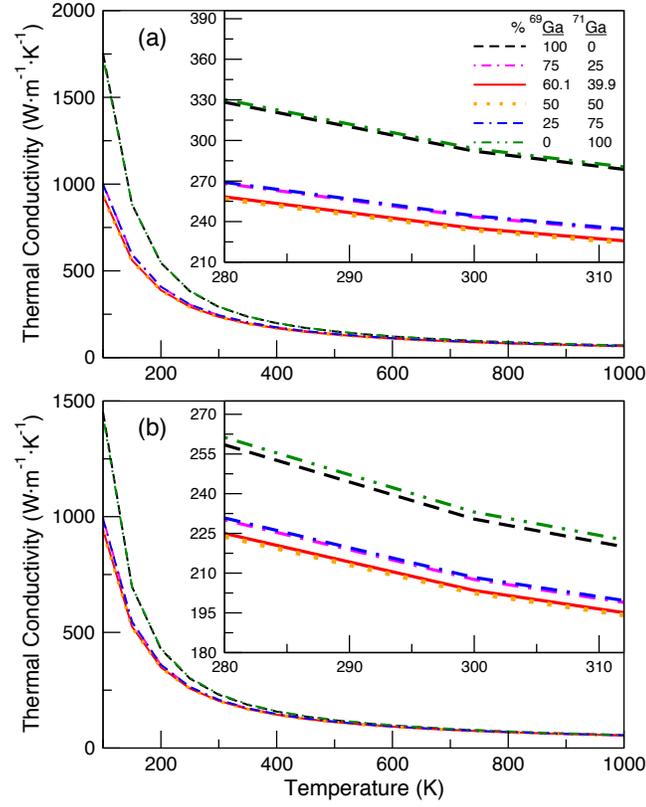

FIG 2. Thermal conductivity of GaN calculated along the (a) **a** and (b) **c**-directions for different ratios of Ga isotopes over the temperature range from 100 to 1000 K. The insets highlight the region near room temperature (280 to 310 K) to illustrate the isotope effects.

TABLE III. Thermal conductivity values of GaN at 300 K (in W m$^{-1}$ K$^{-1}$) along the **a** and **c** crystallographic directions of the wurtzite structure as a function of Ga isotope substitution.

| % $^{69}$Ga | **a**-direction | **c**-direction |
|---|---|---|
| 100 | 292 | 230 |
| 75 | 243 | 208 |
| 60.1 (natural) | 235 | 203 |
| 50 | 234 | 202 |
| 25 | 244 | 208 |
| 0 | 294 | 233 |

To determine the maximum isotope enrichment of thermal conductivity of GaN, we calculated the thermal conductivity with both pure $^{71}$Ga and pure $^{14}$N isotopes (Fig. S1). Using a combination of pure N/pure Ga isotopes increases the thermal conductivity by only 0.2% compared to natural N/pure Ga isotopes, showing that Ga dominates isotope-disorder scattering in the thermal conductivity of GaN.

We also calculated the thermal conductivity of Al$_x$Ga$_{1-x}$N alloys along **a** and **c** within the VCA from 100 to 1000 K and all compositions (Fig. 3). The dependence on composition shows the expected dramatic decrease in $\kappa$ with the introduction of compositional disorder rather than a straight line between the two endpoints. Intermediate compositions fluctuate slightly from a smooth curve due to numerical noise. Furthermore, we fit our calculated thermal-conductivity data using a mathematical model to aid quick and simple evaluations of the phonon-limited thermal conductivity as a function of composition and temperature, $\kappa(x,T)$. The thermal resistivity $\rho(x,T) = 1/\kappa(x,T)$ is given as a function of composition $x$ and temperature $T$ by:

$$\rho(x,T) = \rho_{\text{average}}(x,T) + \rho_{\text{alloy}}(x,T), \tag{1}$$

where $\rho_{\text{average}}$ is the linear average of the thermal resistivity of GaN and AlN (Vegard's law):

$$\rho_{\text{average}}(x,T) = x\rho_{\text{AlN}}(T) + (1-x)\rho_{\text{GaN}}(T), \tag{2}$$

while the thermal resistivity of the binaries is fitted by:

$$\rho_{\text{AlN}}(T) = \rho_{0,\text{AlN}} \times T\left(1 - e^{-\frac{T}{T_0}}\right) \tag{3}$$

and similarly for $\rho_{\text{GaN}}(T)$. The alloy-disorder term ($\rho_{\text{alloy}}$) is given by:

$$\rho_{\text{alloy}}(x,T) = \frac{f(T)x^{m(T)} + g(T)(1-x)^{m(T)}}{(x(1-x))^{n(T)}}, \tag{4}$$

where $f(T)$, $g(T)$, $m(T)$, and $n(T)$ are polynomial functions of T whose fitted coefficients are found in Tables SI-SIII.

Figure 3 shows the thermal conductivity of Al$_x$Ga$_{1-x}$N along **a** and **c**, including both the explicit calculations and the fitted model, over all compositions and temperatures from 100 to 900 K. The fits are in overall excellent agreement with the calculated data and provide a mathematical model to evaluate the thermal conductivity of $Al_xGa_{1-x}N$ alloys for arbitrary composition and temperature. The model fit agrees with the raw calculated data within 5% along **a** and within 10% along **c** across the entire computed temperature and composition range. For all alloy compositions, the thermal conductivity is lower along **a** than along **c**, which is the opposite trend than the binaries. The thermal conductivity decreases with increasing temperature as scattering due to thermal effects increases. Increasing the Al mole fraction also decreases $\kappa$ for $Al_xGa_{1-x}N$ compositions for $x < 0.60$. Our calculations predict the lowest $\kappa$ to occur for Al mole fractions between $x$ = 0.60 - 0.71 in both crystallographic directions over the calculated temperature range.

Figure S2 focuses on the thermal conductivity of $Al_xGa_{1-x}N$ at compositions near the end compounds. Introducing a small amount of compositional disorder decreases the thermal conductivity drastically. The addition of Ga into AlN induces a stronger reduction of the thermal conductivity than a proportionate addition of Al into GaN. For example, at 300 K, adding 1% Al into GaN decreases $\kappa$ by 46.5%, while adding 1% Ga into AlN decreases $\kappa$ by 75.8%. We attribute this asymmetry to Ga being heavier than Al, thus decreasing $\kappa$ more drastically.

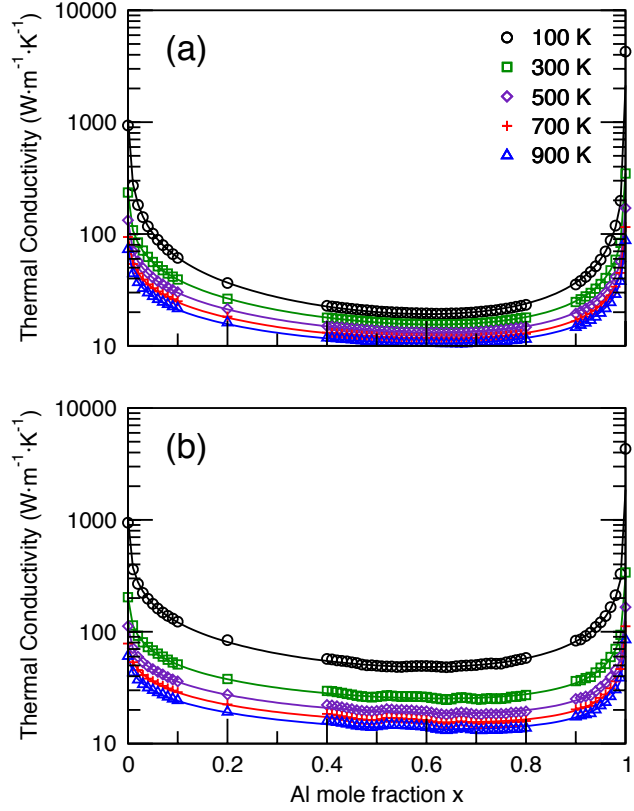

FIG 3. Thermal conductivity of $Al_xGa_{1-x}N$ along the (a) **a**-direction and (b) **c**-direction of wurtzite as a function of Al mole fraction $x$ across the 100-900 K temperature range. The symbols indicate our explicitly calculated data, while the lines indicate the mathematical model [Equations (1-6)] we generated to fit and interpolate the theoretical data.

In summary, we performed first-principles calculations based on the full BTE to study the phonon-limited thermal conductivity of AlN, GaN, and $Al_xGa_{1-x}N$ as a function of temperature, composition, crystallographic direction, and isotope disorder. While using pure $^{14}N$ leads to a 0.2% increase in thermal conductivity in AlN, using pure $^{71}Ga$ and $^{14}N$ can increase the thermal conductivity of GaN by 24% along **a** and 12% along **c**. Our calculations for $Al_xGa_{1-x}N$ span the entire composition range, and we introduce a mathematical model to enable simple calculations of its thermal conductivity at arbitrary composition and temperature. Our study demonstrates the importance of alloy and isotope disorder in the phonon-limited thermal conductivity of AlN,

GaN, and Al$_x$Ga$_{1-x}$N, and our analytical model can guide the thermal design of Al$_x$Ga$_{1-x}$N-based electronic devices.

See supplementary information for figures describing thermal conductivity of GaN with pure Ga and N isotopes and Al$_x$Ga$_{1-x}$N thermal conductivity for edge compositions, as well as polynomial terms and fitting parameters for the alloy-disorder term in the Al$_x$Ga$_{1-x}$N mathematical model.

We thank Jesús Carrete, Natalio Mingo, and Ramon Collazo for helpful discussions on this work. This work was supported by NSF DMREF program (1534221). Computational resources provided by DOE NERSC (DE-AC02-05CH11231). K.A.M. acknowledges the support from the NSF Graduate Research Fellowship Program through Grant No. DGE 1256260.

Irrgeher, R.D. Loss, T. Walczyk, and T. Prohaska, Pure Appl. Chem. **88**, 265 (2016).

# Supplementary Information: Thermal conductivity of AlN, GaN, and Al$_x$Ga$_{1-x}$N alloys as a function of composition, temperature, crystallographic direction, and isotope disorder from first principles

Sahil Dagli, Kelsey A. Mengle, and Emmanouil Kioupakis [a]

Department of Materials Science and Engineering, University of Michigan, Ann Arbor, Michigan 48109,

USA

Figure S1 shows the thermal conductivity of GaN with pure $^{71}$Ga and $^{14}$N in comparison to GaN with naturally occurring isotope ratios.

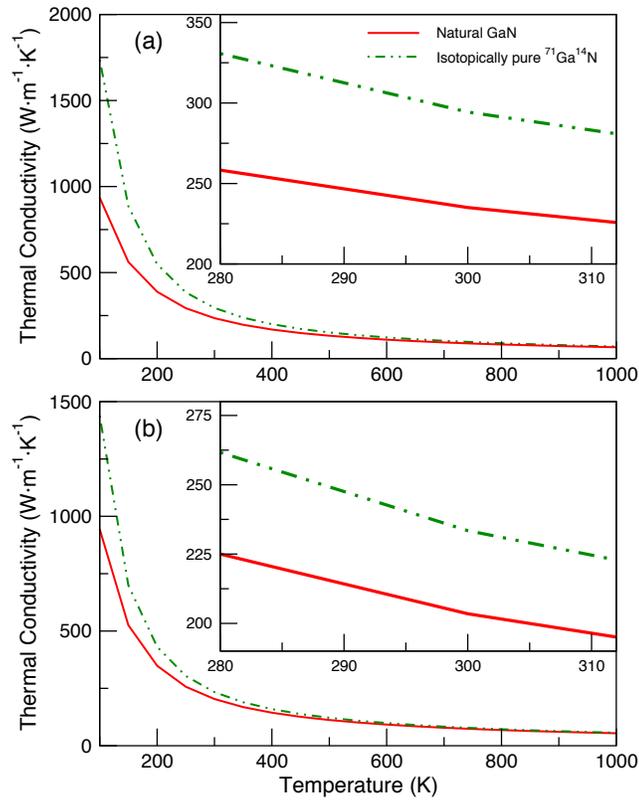

FIG S1. Thermal conductivity of GaN calculated along the (a) **a** and (b) **c**-directions comparing natural GaN with isotopically pure $^{71}$Ga$^{14}$N over the temperature range from 100 to 1000 K. The insets highlight the region near room temperature (280 to 310 K) to illustrate the isotope effects.

In the Al$_x$Ga$_{1-x}$N model, the alloy disorder term contains the following polynomial expressions.

For these terms, the subscript indicates the crystallographic direction. The $f_a(T)$, $g_a(T)$, $m_a(T)$, $n_a(T)$ and $n_c(T)$ terms are given by:

$$f, g, m, n(T) = AT^4 + BT^3 + CT^2 + DT + E, \tag{5}$$

while the $f_c(T), g_c(T),$ and $m_c(T)$ terms are given by:

$$f, g, m(T) = PT^q + RT + S. \tag{6}$$

Tables SI-SIII include the fitting parameters for these expressions.

TABLE SI. Fitted parameters for the $\rho_{average}$ term in our Al$_x$Ga$_{1-x}$N alloy thermal conductivity model [Eq. (3)].

|  | AlN | GaN |
|---|---|---|
| $\rho_{0,a}$ ($m\,W^{-1}$) | 1.29×10$^{-5}$ | 1.52×10$^{-5}$ |
| $\rho_{0,c}$ ($m\,W^{-1}$) | 1.46×10$^{-5}$ | 1.83×10$^{-5}$ |
| $T_{0,a}$ ($K$) | 221 | 105 |
| $T_{0,c}$ ($K$) | 219 | 133 |

TABLE SII. Fitted parameters for the $f_a(T), g_a(T), m_a(T), n_a(T),$ and $n_c(T)$ terms in our Al$_x$Ga$_{1-x}$N alloy thermal conductivity model [Eqs. (4-5)] as a function of composition and temperature (K). Units are indicated in parentheses.

| Coefficient | $f(T)$ ($m\,K\,W^{-1}$) | $g(T)$ ($m\,K\,W^{-1}$) | $m(T)$ (−) | $n_a(T)$ (−) | $n_c(T)$ (−) |
|---|---|---|---|---|---|
| A | −1.21×10$^{-11}$ | −2.02×10$^{-11}$ | 0 | 1.06×10$^{-12}$ | 0 |
| B | 3.39×10$^{-8}$ | 5.89×10$^{-8}$ | 2.73×10$^{-10}$ | −3.15×10$^{-9}$ | −1.35×10$^{-10}$ |
| C | −3.39×10$^{-5}$ | −5.92×10$^{-5}$ | −5.97×10$^{-7}$ | 3.53×10$^{-6}$ | 3.56×10$^{-7}$ |
| D | 9.66×10$^{-3}$ | 1.70×10$^{-2}$ | 7.12×10$^{-5}$ | −1.75×10$^{-3}$ | −2.71×10$^{-4}$ |
| E | 5.74 | 10.7 | 1.35 | 0.896 | 0.639 |

TABLE SIII. Fitted parameters for the $f_c(T), g_c(T),$ and $m_c(T)$ terms in our Al$_x$Ga$_{1-x}$N alloy thermal conductivity model [Eqs. (4) and (6)] as a function of composition and temperature.

| Coefficient | $f(T)$ ($m\,K\,W^{-1}$) | $g(T)$ ($m\,K\,W^{-1}$) | $m(T)$ (−) |
|---|---|---|---|
| P | 2620 | 10380 | 2.55×10$^6$ |
| q | −1.10 | −1.35 | −3.46 |
| R | −1.19×10$^{-3}$ | −5.76×10$^{-3}$ | −1.47×10$^{-4}$ |
| S | 4.17 | 14.1 | 1.03 |

Figure S2 focuses on the thermal conductivity of $Al_xGa_{1-x}N$ at compositions near the AlN and GaN end compounds.

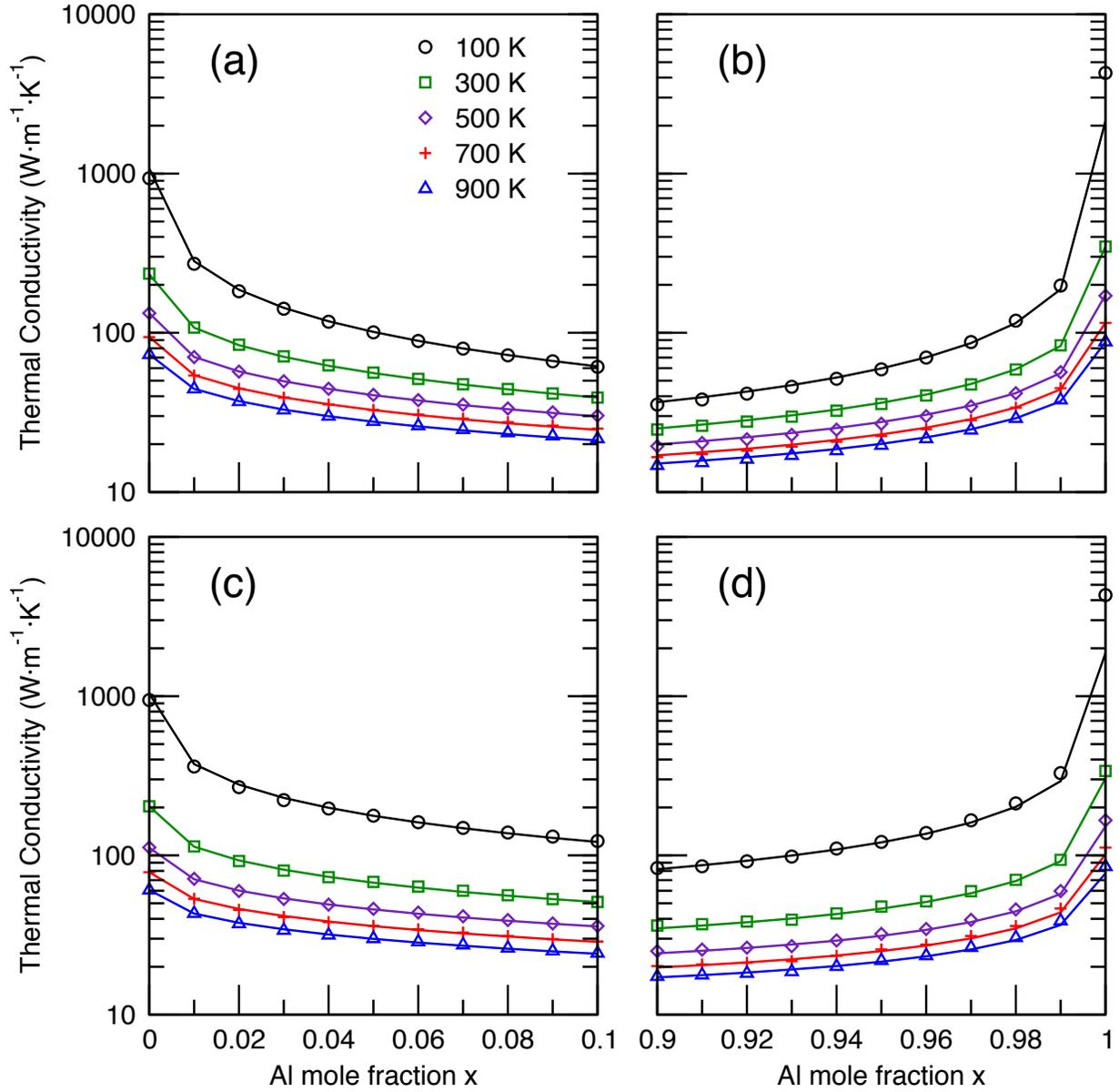

FIG S2. Thermal conductivity of $Al_xGa_{1-x}N$ for compositions near pure GaN in the (a) **a** and (c) **c**-directions, and near pure AlN in the (b) **a** and (d) **c**-directions across the temperature range 100 – 900 K. The symbols indicate our explicitly calculated data, while the lines indicate the mathematical model [Equations (1-6)] we generated to fit and interpolate the theoretical data.